\documentclass[pre,rsi,amsmath,amssymb,reprint]{revtex4-1}

\usepackage{graphicx}
\usepackage{dcolumn}
\usepackage{bm}
\usepackage{natbib}
\setcitestyle{square}
\usepackage{soul}

\usepackage{amsmath}
\usepackage{braket}
\usepackage{mathrsfs}
\usepackage{amsfonts}
\usepackage{color}
\usepackage[normalem]{ulem} 
\usepackage{xcolor}
\usepackage{hyperref}
\usepackage{cleveref}
\DeclareMathOperator{\Tr}{tr}

\usepackage{dirtytalk}

\hypersetup{
    colorlinks=true,
    linkcolor=blue,
    filecolor=magenta,      
    urlcolor=black,
    citecolor=red,
}

\begin{document}

\title{Generalized Naming Game and Bayesian Naming Game as Dynamical Systems}

\author{Gionni Marchetti}
\email{gionnimarchetti@ub.edu}
\email{gionnimarchetti@gmail.com}


\affiliation{%
Departament de F\'{i}sica de la Mat\`{e}ria Condensada, Facultat de Física,  Universitat de Barcelona, Carrer Mart\'{i} i Franqu\`{e}s 1, 08028, Barcelona, Spain\\
Institut de Nanoci{\`e}ncia i Nanotecnologia, Universitat de Barcelona, Av. Joan XXIII S/N, 08028, Barcelona, Spain\\
}%

\date{\today}

\begin{abstract}
We study the $\beta$-model ($\beta$-NG) and the Bayesian Naming Game (BNG) as dynamical systems. By applying linear stability analysis to the dynamical system associated with the $\beta$-model, we demonstrate the existence of a non-generic bifurcation with a bifurcation point $\beta_c = 1/3$. As $\beta$ passes through $\beta_c$, the stability of isolated fixed points changes, giving rise to a one-dimensional manifold of fixed points. Notably, this attracting invariant manifold forms an arc of an ellipse. In the context of the BNG, we propose modeling the Bayesian learning probabilities $p_A$ and $p_B$ as logistic functions. This modelling approach allows us to establish the existence of fixed points without relying on the overly strong assumption that $p_A = p_B = p$, where $p$ is a constant.
\end{abstract}

\maketitle

\section{Introduction} \label{introduction}

The  naming game is a multi-agent model that aims to simulate the spontaneous emergence of consensus among agents who interact pairwise  according to two possible processes: agreement and word/name learning~\cite{Baronchelli_2006, Blythe2015, Baronchelli-2016a, chen-2019a, Marchetti-2020b}. In its simplest realization, the agents are allowed to learn only two possible names, denoted here by the symbols A and B. Furthermore, if the  underlying topology on which the agents may lie, is  neglected altogether~\cite{Castello-2009a, Marchetti-2020b}, we refer to such a model as the minimal naming game (MNG).

 This multi-agent model can be thought as a particular case of two variants: the generalized naming game ($\beta$-NG)~\cite{baronchelli2007Main} and the Bayesian naming game (BNG)~\cite{Marchetti-2020a, Marchetti-2020b, MARCHETTI2021}. Both variants add human features otherwise absent in agents of  MNG. In $\beta$-NG, a parameter $\beta \in [0, 1]$ is introduced to take into account the probability of acknowledged influence~\cite{Noah1990}, thus affecting the agreement processes, while in BNG the name learning processes are now dictated by the time-dependent probabilities $p_A \equiv p_A\left(t\right)$ and $p_B \equiv p_B\left(t\right)$ necessary for generalizing the concept $\mathcal{C}$ associated with words A and B, respectively ~\cite{Tenenbaum-1999,  Tenenbaum2001, Murphy-2012a}. It is worth noting here that such probabilities are a direct consequence of  Bayes rule~\cite{Jeffreys1961}. Therefore, according to the latter model the agents become Bayesian-rational individuals that in an uncertain world find their best action using the laws of probability~\cite{deFinetti1989, auman1976, Aaronson2005, Sanborn2016}. We refer the reader to Section~\ref{naming_games} for a detailed description of MNG and its variants $\beta$-NG and BNG.

The  dynamics of the naming game is routinely investigated through computer simulations of multi-agent systems. The simulations allow for the computation of the observables of interest, e.g. the convergence time, the success rate, averaging over several hundred realizations~\cite{Marchetti-2020b}. However,  the differential equations dictating the time-evolution for each model, can be also derived, assuming the mean-field approximation, thus offering an alternative approach to study some features of consensus dynamics. In fact, the differential equations of these models form two-dimensional (2D) dynamical systems whose study has some evident advantages.  First,  due to their low dimensionality, the time-evolution is much simpler when compared to that of naming game models where multiple names need to be taken into account~\cite{chengMa2023}. Second, one can exploit the methods provided by the dynamical systems theory~\cite{strogatz2018, hirsch2013}, such as bifurcation diagrams and phase portraits for autonomous systems, to infer whether or not the consensus emerges  without the need to perform the numerical simulations, a task that becomes computationally expensive when  the system size $N$, i.e., the number of agents, increases.

In this paper, we shall pursue the latter approach, thereby finding novel interesting dynamical features of the dynamical systems associated with the $\beta$-NG and BNG, which, to our best knowledge, have been overlooked so far. 

To begin with, by systematically applying the linear stability (LSA) to the dynamical system associated with $\beta$-NG we show that such a  dynamical system undergoes to a non-generic bifurcation with bifurcation point  $\beta =\beta_c$ where $\beta_c=1/3$ (see Section~\ref{dynamics_beta}). This is an important finding because it shows that the first-order (nonequilibrium) phase transition at $\beta = \beta_c$ obtained by means of numerical simulations in Ref.~\cite{baronchelli2007Main} corresponds to a special feature in the dynamical system approach: a bifurcation
where several  things happen simultaneously. In fact, there is an emergence of 
one-dimensional manifold of fixed points forming an arc of an ellipse, when $\beta$ goes through $\beta_c$. 
Additionally, this bifurcation  affects the type of  noise
 present in the trajectories from multi-agent simulations, as they are interpreted as sample paths of two important stochastic processes: the Brownian motion and  Ornstein-Uhlenbeck (OU) process~\cite{erban2009, Oksendal2010}.

Regarding the dynamical system associated with the BNG, in Section~\ref{dynamics_bayes}, we shall show that the overly strong assumption that $p_A = p_B = p$, where $p$ is a constant, cannot be used to prove the existence of fixed points and, hence, the existence of consensus, as proposed in Ref.~\cite{Marchetti-2020a}, in full generality, due to the lack of a spontaneous symmetry breaking process.
To address this serious problem, we propose modeling the Bayesian probabilities as time-dependent logistic functions~\cite{james2023}. Subsequently, by numerically integrating the corresponding differential system, it is shown that there exist two fixed points and also that its solution closely resembles that obtained from multi-agent simulations with a system size $N \gtrsim 200$~\cite{Marchetti-2020a}.

\section{Minimal Naming Game and its variants:  $\beta$-NG  and BNG} \label{naming_games}

According to the BNG model, at each time-step, two agents are randomly chosen within a system of size $N$. Then, the first agent, who acts as a speaker, selects the name stored in her inventory or randomly chooses one of the two possible names A and B if both are therein stored, and then utters it to the second agent (the hearer)~\footnote{In our simulations, the agents do not need to invent the names from scratch, as one name, either A or B, is assigned to each agent's inventory at the beginning.}. 
If the inventory of the hearer contains such a name, then both agents update their inventories, keeping only the conveyed name (agreement). Otherwise, the hearer learns the new name and subsequently adds it to her inventory (deterministic word learning, as it occurs with a probability equal to unity).~\cite{Baronchelli_2006, chen-2019a}. Note that through the above pairwise interactions, the agents attempt to reach a consensus on the name to assign to a single object~\cite{Baronchelli-2016a}. Such a consensus can be achieved when all the agents have the same single name in their inventories. 

According to the $\beta$-NG model, the previous agreement process requires a modification. Specifically, if both agents have the communicated name in their inventories, they are allowed to update their inventories with a probability of $\beta$, while nothing changes with a probability of $1-\beta$~\cite{baronchelli2007Main}.

On the other hand, the BNG model replaces the above deterministic learning process with a human-like word learning process under uncertainty based on Bayesian inference~\cite{Tenenbaum-1999,  Tenenbaum-1999b, Murphy-2012a, Griffiths2006, Griffiths-2007, Tenenbaum-2011a, Perfors-2011a, Lake2015}, while leaving the agreement process unchanged. The key idea underpinning this model is that learning a word requires multiple cognitive efforts over an extended period of time, and also requires the learning (or generalization) of the concept $\mathcal{C}$ associated with it from a few positive examples (few-shot learning). In this regard, the BNG model paves the way for incorporating many human cognitive biases~\cite{Tversky1974,  kahneman2012, Hahn2014, Sanborn2016, Ngam2016, Madsen2018} into the consensus dynamics, which were out of reach for the previous variants of naming game. 

Accordingly, such a probabilistic model requires that the speaker not only utters a name but also, at the same time, provides a positive example associated with $\mathcal{C}$, randomly chosen from his or her inventory.~\footnote{Therefore, the agents are now equipped with three inventories: one for storing the names A and/or B, and two for storing the positive examples associated with A and B.}. For computational convenience, these positive examples are represented as points belonging to an axis-parallel rectangle, i.e., a rectangle concept, in $\mathbb{R}^{2}$~\footnote{In our case the rectangle is defined the following tuple $\left( x, y,  \sigma_1,  \sigma_2 \right)$  where $x=0, y=0$ are the lower-left corner's  coordinates and  $\sigma_1=3,  \sigma_2=1$ the rectangle's sizes along the $x$ and $y$-axis, respectively.}. Next, the hearer needs to compute the Bayesian probability for generalizing $\mathcal{C}$ from the conveyed example together with those previously stored, assuming an Erlang prior~\cite{Tenenbaum-1999, Marchetti-2020a, MARCHETTI2021}. The learning process is successful whenever the value of such a probability is equal to or greater than a threshold value, which we set to $0.5$~\cite{Marchetti-2020a}.

 In present work, all the multi-agent simulations will start with initial condition corresponding to  $50\%$  of the agents knowing the word A and the remainder knowing the other word B~\cite{Marchetti-2020a}. Note that from a dynamical system point of view, this setup corresponds exactly to having the initial condition $I_0=\left(0.5, 0.5\right)$ in the phase plane (see Fig.~\ref{fig:phasePortrait}).
 
Furthermore, at the beginning of the simulation we randomly  assigned an number of four positive examples to each agent's  inventory.  However, there are different  thresholds $n^{\ast}_{ex, A}=5$,  and $n^{\ast}_{ex, B}=6$, of the number of positive examples stored by the agents, necessary for computing the probabilities for generalizing the concept, respectively. As a consequence of the last assumption, the words A and B are now distinguishable synonyms, and, at the same time, the learning of A is favored~\cite{Marchetti-2020a}.

Finally, tt is worth noting here that in MNG and BNG models the consensus always  emerges spontaneously, but the same can occurs only when
$\beta>\beta_c$ in  $\beta$-model. Furthermore,  the MNG model can be recovered from its two variants  by setting their characteristic parameters, i.e. $\beta$ and $p_A, p_B$, to unity.

\section{The $\beta$-NG model as Dynamical System}\label{dynamics_beta}

According to $\beta$-NG model, in  mean-field approximation the time-evolution of the population fractions $x\equiv x\left(t\right)$ and $y\equiv y\left(t\right)$, having either A or B in their inventories at time $t$ respectively, is dictated by the following system of  first-order differential equations~\cite{Castello-2009a}:
\begin{align}
\label{eq:dot1}
&\dot{x} = - x y + \beta \left(1 - x- y \right)^{2} +  \frac{3 \beta -1 }{2} x \left(1 - x- y \right)  \, ,
\\
\label{eq:dot2}
&\dot{y} =  - x y + \beta \left(1 - x- y \right)^{2} + \frac{3 \beta -1 }{2} y \left(1 - x- y \right)   \, ,
\end{align}
where $\dot{x}, \dot{y}$ the derivatives of $x,y$, with respect to time. Note that hereafter, for convenience, we will utilize the symbol $K\equiv \left(3\beta -1\right)/2$.

Next, setting $\dot{x} = 0$ and $\dot{y} = 0$ into Eqs.\ref{eq:dot1} and \ref{eq:dot2}, it was previously determined that three isolated fixed points exist: $A=\left(1, 0\right)$,  $B=\left(0, 1\right)$  and $C=\left(c_{\beta}, c_{\beta}\right)$ whose  coordinates in terms of parameter $\beta$ are given by $c_{\beta} =  \left( 5 \beta + 1 - \sqrt{\Delta } \right)/ 4 \beta$ where $\Delta \equiv 1 +  10 \beta + 9 \beta^2$ ~\cite{baronchelli2007Main, castellano2009}~\footnote{Note that  it was originally found that $\Delta = 1 +  10 \beta + 17 \beta^2$.}. Additionally, the LST revealed a first-order non-equilibrium transition at $\beta = \beta_c$, but the corresponding bifurcation went unnoticed. In such a scenario, the system converges to point C for $\beta < \beta_c$, leaving a fraction of undecided agents equal to $1 -2c_{\beta}$, while a spontaneous emergence of consensus occurs at either A or B for $\beta > \beta_c$. These findings are summarized in the first and third columns of Table~\ref{table:table1}.

While the above analysis is certainly correct and has been confirmed by multi-agent simulations, it does not address the non-generic bifurcation that characterizes the dynamical system in question.

In the following, we shall unravel the dynamical features of the dynamical system
 by applying the LSA to the problem at hand in a systematic way. First, let us define the functions  $f_1\left(x,y\right), f_2\left(x,y\right)$ such that $\dot{x} = f_1\left(x,y\right), \dot{y} = f_2\left(x,y\right)$ according to  Eqs.~\ref{eq:dot1}, \ref{eq:dot2}. It is then possible to compute the eigenvalues of  the following
 Jacobian matrix $J$ evaluated at a given fixed equilibrium point with Cartesian coordinates $\left(\Tilde{x}, \Tilde{y}\right)$~\cite{strogatz2018}
\begin{equation}\label{eq:jacobian1}
J =
\begin{pmatrix}
  \partial_x f_1\left(\Tilde{x}, \Tilde{y}\right)  &   \partial_y f_1\left(\Tilde{x}, \Tilde{y}\right)  \\
 \partial_x f_2\left(\Tilde{x}, \Tilde{y}\right)  &   \partial_y f_2\left(\Tilde{x}, \Tilde{y}\right)    
\end{pmatrix} \, ,
\end{equation}
where the symbols  $\partial_x $,  $\partial_y$ denote the partial derivatives of  $f_1\left(x, y\right)$,  $f_2\left(x, y\right)$ with respect to $x, y$, respectively. 

\begin{figure}

\resizebox{0.50\textwidth}{!}{%
  \includegraphics{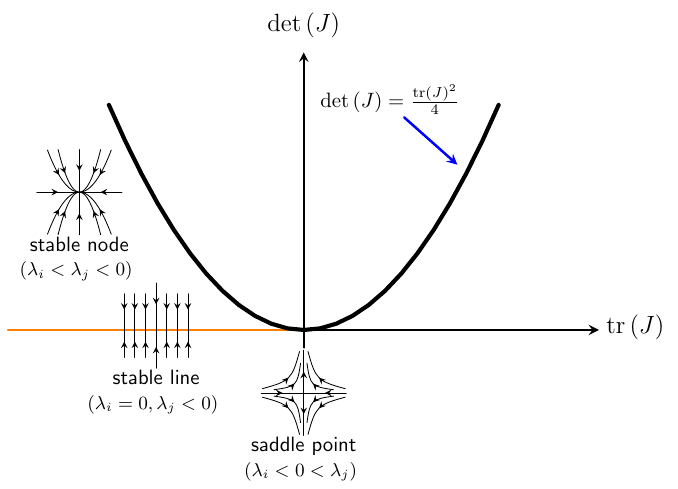}
}

\caption{The types of the equilibria for the dynamical system with Eqs.~\ref{eq:dot1}, \ref{eq:dot2} and $\beta \in [0,1]$  according to
LSA, in the trace-determinant plane. 
The respective inequalities between the two (real)  eigenvalues $\lambda_i, \lambda_j$ ($i \neq j$) of the  matrix $J$ evaluated at each type of fixed point  are also shown.
The parabola in the (black) solid line corresponds to  $ \Tr \left(J\right)^{2} - 4 \det \left(J\right)   = 0$. Note that all the saddle points lie on quadrants III and IV of the  trace-determinant plane.}
\label{fig:bifurcationDiagram}     
\end{figure}

Second, the local equilibrium character of each fixed point can be determined by the eigenvalues of the corresponding matrix $J$. In particular, their equilibrium character is easily identified by looking at the inequalities between the eigenvalues as shown in the trace-determinant plane (also called the bifurcation diagram)  (see Fig.~\ref{fig:bifurcationDiagram}) ~\cite{hirsch2013,  Perez-Garcia2019}).

\begin{figure}

\resizebox{0.50\textwidth}{!}{%
  \includegraphics{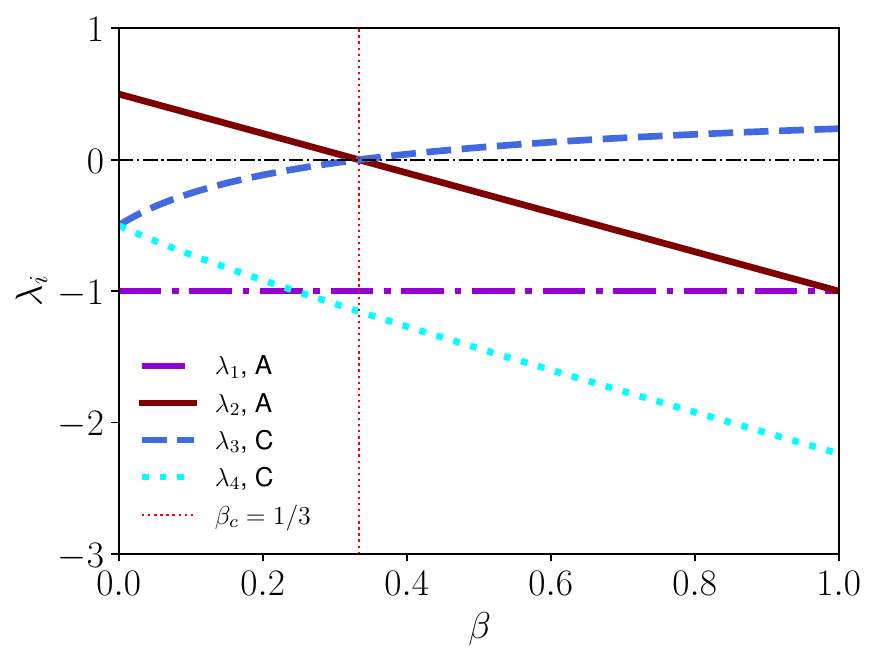}
}

\caption{The real eigenvalues  $\lambda_i$  with $i= 1,2$ and   $i= 3,4$ corresponding to the Jacobian matrix J (Eq.~\ref{eq:jacobian1}) evaluated at  A and $C$, respectively, as functions of the parameter $\beta$.  The vertical (dotted) line corresponds to $\beta=\beta_c$. } 
\label{fig:eigenvalues}     
\end{figure}

The  matrix $J$'s entries evaluated at the fixed point A read: $J_{11} = -K$, $J_{12} = -1 -K$, $J_{21} = 0$, $J_{22} = -1$ while the matrix entries takes the values
 $J_{11} = -1$, $J_{12} = 0$, $J_{21} = -1-K$, $J_{22} = -K$ when computed at B. Since in both cases the traces and the determinants of the Jacobian matrices are the same, i.e. $\Tr \left(J\right)= -1 - K$ and $ \det \left(J\right) =K $,  one must conclude that these matrices have same eigenvalues. Therefore, the fixed points $A,B$ present the very same stability behaviour for each value of  $\beta$ in $[0, 1]$. Accordingly, in the following we shall focus on the equilibrium character of A only, denoting its  corresponding eigenvalues by the symbols $\lambda_1, \lambda_2$. By means of the equations  $\Tr \left(J\right)= \lambda_1 +  \lambda_2 $ and  $ \det \left(J\right) = \lambda_1 \lambda_2$, one immediately finds these two real eigenvalues: $\lambda_1 = -1$ and $\lambda_2 = -K$. Since $\lambda_1$  is always negative, both eigenvalues are negative when  $\beta > \beta_c$, while  $\lambda_2>0$ when $\beta < \beta_c$ (see Fig.~\ref{fig:eigenvalues}). Looking at  the classification scheme  in  Fig.~\ref{fig:bifurcationDiagram}, one must conclude that $A, B$ are saddle points and asymptotically stable nodes for  $\beta \in  [0, \beta_c)$ and $\beta \in  (\beta_c, 1] $, respectively. So far, the results agree with those reported in Ref.~\cite{baronchelli2007Main}. However, 
 at $\beta =\beta_c$, $\lambda_2 = 0$, and hence both A and B must belong to the stable line corresponding to the negative half-axis (see the (orange) solid line in Fig.~\ref{fig:bifurcationDiagram}). Therefore, A and B cannot be longer considered isolated, but according to LSA there must exist an infinite number of other fixed points. We will return to this point later.

We now turn our attention to the equilibrium character of the fixed point $C$, denoting its corresponding eigenvalues by $\lambda_3$ and  $\lambda_4$. To this end, we shall compute them  by means of the following formula ~\cite{sagues2003}

\begin{equation}\label{eq:eigenvalues}
\lambda_3, \lambda_4 = \frac{1}{2} \left(\Tr \left(J\right)  \pm \sqrt{\Tr \left(J\right)^{2} - 4 \det \left(J\right)}\right) \, .
\end{equation}

In Fig.~\ref{fig:eigenvalues}  we plot  $\lambda_3, \lambda_4$ as function of $\beta$ by means of Eq..~\ref{eq:eigenvalues}. From a direct comparison of the corresponding curves with those of $\lambda_1, \lambda_2$, one immediately concludes that $C$ has the opposite equilibrium character of A and B for each each $\beta \in [0, 1] \setminus \{\beta_c\} $, see Fig.~\ref{fig:eigenvalues}. 
However, when $\beta =\beta_c$, $C$  belongs to the stable line as well because  $\lambda_3 =0$  while  $\lambda_4$ negative (see Fig.~\ref{fig:bifurcationDiagram}).

Noting that the stable line is characterized by $ \Tr \left(J\right)<0$ and $\det \left(J\right)   = 0$, the last important result could also be obtained without the need to compute the eigenvalues, as follows.
First, we note that at  $C=\left(c_{\beta}, c_{\beta}\right)$,  the  matrix $J$'s entries must satisfy the identities $J_{11} = J_{22} $ and   $J_{12} = J_{21} $, thus the Jacobian matrix is now symmetric. Second, $J_{11}, J_{12}$ take the following form as functions of $\beta$
\begin{equation}\label{eq:matrixElement1}
J_{11} = - \left(\frac{\beta +1}{2}\right) + \left(\frac{1 -\beta }{2}\right) c_{\rm \beta}\, ,
\end{equation}
and 
\begin{equation}\label{eq:matrixElement2}
J_{12} = - 2\beta  + \left(\frac{5\beta -1}{2}\right)c_{\rm \beta} \, ,
\end{equation}
respectively.

Therefore, in such a case $\Tr \left(J\right) = 2 J_{11}$ and  $ \det \left(J\right)=  J_{11}^{2} - J_{12}^{2}$. Now noting that $ \det \left(J\right)= \left(J_{11} + J_{12}\right)  
\left(J_{11} - J_{12}\right)$, one finds that 
\begin{equation}\label{eq:matrixElement2}
J_{11} - J_{12} = K \left(1 - 2  c_{\beta} \right)  \, .
\end{equation} 

So, as  $K =0$ when $\beta=\beta_c$, it follows that the left hand side of Eq.~\ref{eq:matrixElement2}, and hence $\det \left(J\right)$ are both zero. Next,  we note that the corresponding  trace is  negative, i.e. $\Tr \left(J\right) = 2/3 \left( c_{\beta} -2 \right) <0$, because  $c_{\beta} = 2 - \sqrt{3} \approx 0.26$. 

Overall, our analysis shows that the fixed points $A, B, C$ exist for all values of the parameter $\beta$ but there an exchange of stabilities between $A, B$ and $C$ as $\beta$ passes through $\beta_c$. In this regard, we can say that the dynamical system undergoes a non-generic bifurcation similar to that observed in the 1D differential equation
$\dot{z} = \xi \left(z - z^{3}\right)$ at $\xi=0$ ($\xi$ is a parameter) with fixed isolated points $0, \pm 1$. In fact, as $\xi$ passes through $0$, the fixed points at $z = 0$, $z = 1$, and $z = –1$ change stability and a line of fixed points emerges (the whole  $z$-axis is filled with fixed points when $\xi = 0$). This is a very degenerate situation because the single parameter $\xi$ affects both the linear and the cubic term in the differential equation simultaneously, causing both terms to vanish when $\xi = 0$. For the sake of completeness, we summarize all these findings in Table~\ref{table:table1}.

\begin{table}

\renewcommand{\arraystretch}{2} 
    \caption{The classification of the fixed points $A, B, C$ according to the linear stability theory. Note that when the isolated equilibrium points are stable, they are also asymptotically stable. The fixed points $A, B, C$ are no longer isolated at bifurcation point $\beta=\beta_c$.}
\label{table:table1}
\begin{ruledtabular}
\begin{tabular}{c c c c   }
Fixed Point & $\beta \in [0,\beta_c) $ & $\beta= \beta_c$ &  $\beta \in (\beta_c, 1]$  \\ [1ex]
\hline
    A    & saddle point         & stable line    & stable node   \\
    B     & saddle point             & stable line   & stable node      \\
    C     & stable node           & stable line       & saddle point   \\ [1ex]
\end{tabular}
\end{ruledtabular}
\end{table}

\begin{figure}
\resizebox{0.45\textwidth}{!}{%
  \includegraphics{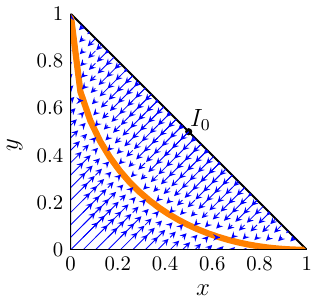}
}

\caption{Streamline plot  of   $\mathbf{v} = \left(\dot{x}, \dot{y} \right)$  corresponding to Eqs.~\ref{eq:dot1}, \ref{eq:dot2} at bifurcation point $\beta=\beta_c$.   Here the curve (orange) of the attracting invariant manifold corresponds to the graph of the map  $x \mapsto  1/2 \left(-\sqrt{3}\sqrt{4x -x^{2}}   + x + 2 \right)$ with $x \in [0,1]$.  The important initial condition $I_0 =\left(0.5,0.5\right)$ is also shown. }
\label{fig:phasePortrait}       
\end{figure}

Next, the existence of a continuum of fixed points at the bifurcation point could be easily inferred by examining the streamline plot of the velocity field $\mathbf{v} = \left(\dot{x}, \dot{y} \right)$ corresponding to Eqs.~\ref{eq:dot1}, \ref{eq:dot2}.
To this end, in Fig.~\ref{fig:phasePortrait} we plot such a velocity field on the model's domain in   Cartesian $xy$-plane: 
$\{ (x,y) \in \mathbb{R}^2: 0  \leq   x \leq 1,  0  \leq  y \leq 1, 0 \leq  x + y  \leq 1 \}$.
It is then evident that there exists one-dimensional manifold of fixed points (orange solid line), and that all trajectories approach it along parallel lines. This finding is consistent with the stable line, that is, the negative negative half-axis  of trace $\Tr J$, obtained from LSA (see Fig.~\ref{fig:bifurcationDiagram}). Furthermore, we note that the flow of the above 1D differential equation is similar to the 2D  flow  along the attracting invariant manifold corresponding to the (orange) stable ``line'' in the bifurcation diagram (see Fig.~\ref{fig:bifurcationDiagram}) and to the (orange) curve in the 2D phase plane (see Fig.~\ref{fig:phasePortrait}).

We now prove that the above one-dimensional manifold constitutes an arc of an ellipse. First, its Cartesian equation can be  obtained by setting $\beta=\beta_c$ and either  $\dot{x}=0$ or $\dot{y}=0$ into Eqs.~\ref{eq:dot1}, \ref{eq:dot2}.  One then finds that the set of points must satisfy the following second-order degree polynomial equation in the variables  $x,y$
\begin{equation}\label{eq:quadratic}
x^{2} -xy + y^{2} -2x -2y + 1 = 0 \, .
\end{equation}

\begin{figure}
\resizebox{0.35\textwidth}{!}{%
  \includegraphics{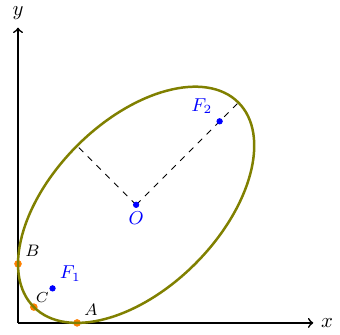}
}

\caption{The graph of the ellipse with Eq.~\ref{eq:quadratic} emerging  at bifurcation point $\beta=\beta_c$. The ellipse's center and foci are $0=\left(2, 2\right)$, $F_1= \left(2-\sqrt{2}, 2-\sqrt{2}\right)$ and  $F_2= \left(2+\sqrt{2}, 2+\sqrt{2}\right)$, respectively. The fixed points $A=\left(1, 0\right)$, $B=\left(0, 1\right)$ and  $C=\left(c_{\beta_c}, c_{\beta_c}\right)$ where $c_{\beta_c}=2 -\sqrt{3}$ are no longer isolated equilibria.}
\label{fig:ellipse}      
\end{figure}

According to Eq.~\ref{eq:quadratic} the one-dimensional manifold corresponds to the graph of the function $x \mapsto  1/2 \left(-\sqrt{3}\sqrt{4x -x^{2}}   + x + 2 \right)$ with $x \in [0,1]$.
Next, we construct the symmetric matrix $M$ of the quadratic form associated with Eq.~\ref{eq:quadratic} in order to classify it as a nondegenerate conic section~\cite{Apostol1969}. In the present case, one finds
\begin{equation}\label{eq:matrix}
M =
\begin{pmatrix}
  1  &   0.5  \\
 0.5 &   1   
\end{pmatrix} \, ,
\end{equation}
with determinant   $\det \left(M\right) = 3/4$. As a consequence its eigenvalues must have the same sign, so Eq.~\ref{eq:quadratic} yield a conic section that is an ellipse with semi-major axis,   semi-minor axis and  eccentricity corresponding to $a=\sqrt{6}$,  $b=\sqrt{2}$ and $e=\sqrt{2/3}$, respectively. The graph of such an ellipse  with its center  $0 =\left(2,2\right)$ and foci    $F_1 = \left(2 -\sqrt{2},2 - \sqrt{2}\right)$ and $F_2 =\left(2 +\sqrt{2},2 + \sqrt{2}\right)$ is shown in  Fig.~\ref{fig:ellipse}. Note that in such a case, point $C$ now has coordinates $\left(c_{\beta_c}, c_{\beta_c}\right)$ with $c_{\beta_c}=2 -\sqrt{3}$ and hence lies on the ellipse between points A and B.

We conclude this section by noting another important consequence of the existence of a non-generic bifurcation. The dynamical system approach provides a deterministic mean-field description of the $\beta$-model, but the trajectories from its multi-agent simulations are intrinsically noisy. Therefore, one would expect that there is some effect of noise close to the bifurcation~\cite{erban2009}.

\begin{figure}
\resizebox{0.50\textwidth}{!}{%
  \includegraphics{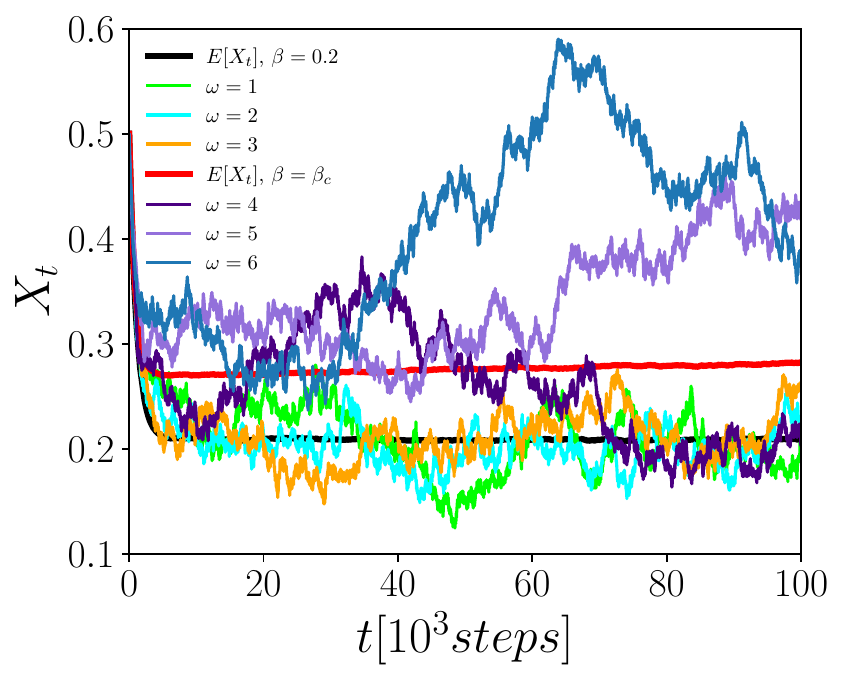}
}

\caption{Randomly chosen sample paths of $X_t$ ($X_0 =0.5$) as functions of time $t$ from multi-agents simulations with $N= 1,000$ agents. The paths with label $\omega=1, 2, 3$ and $\omega=4, 5, 6$ correspond to $\beta= 0.2$ and $\beta=\beta_c$, respectively. Solid lines in black color and red color correspond to sample mean  $E[X_t]$ computed from $1, 200$ trajectories for $\beta=0.2$ and  $\beta=\beta_c$, respectively. }
\label{fig:samplepath}       
\end{figure}

To illustrate this, let us consider $x$, or equivalently $y$, obtained from a single run of a simulation starting from $I_0$ as a random variable $X_t$. In Fig.~\ref{fig:samplepath}, we plot six randomly chosen sample paths of $X_t$ as functions of time from simulations with $N=1,000$ agents. Note that the paths labeled by $\omega=1, 2, 3$ and $\omega=4, 5, 6$ correspond to $\beta=0.2$ and $\beta=\beta_c$, respectively where each $\omega_i$ could be understood as a specific outcome in the sample space.

The paths with $\omega=1, 2, 3$ exhibit a pattern completely different from those at the bifurcation point, i.e., with $\omega=4, 5, 6$. While the former remain close to the sample mean $E[X_t]$ (black solid line) computed from $1, 200$ trajectories, thus showing the typical mean-returning behavior of the  OU process
\cite{Uhlenbeck1930, vasicek1977}, the latter seem to wander in a more random fashion. Preliminary statistical analysis~\cite{gionni2024} confirms that the sample paths at  $\beta= 0.2$ belong to OU process, while  those at bifurcation point   to a (nonstandard) Brownian motion~\cite{einstein1905, langevin1908, wiener1923, inventions4020024} (see Section~\ref{variance} for details).

\section{The Bayesian Naming Game as a Dynamical System}\label{dynamics_bayes}

Assuming the mean-field approximation, the time evolution of BNG is governed by the following non-autonomous first-order differential system~\cite{Marchetti-2020a}:

\begin{align}
  \dot{x} = - p_B x y + \left(1- x - y \right) ^{2} + \frac{3 - p_B}{2} x \left(1- x - y \right)  \, , \label{eq:nonautonomous1}\\  
  \dot{y} = - p_A x y +  \left(1- x - y \right)^{2} +  \frac{3 - p_A}{2} y \left(1- x - y \right) \label{eq:nonautonomous2} \, .
\end{align}

Note that finding a solution to the above system is not possible without making some reasonable assumptions about the probabilities $p_A\left(t\right)$ and $p_B\left(t\right)$, both falling within the interval [0,1] with an unknown functional form. In Ref.\cite{Marchetti-2020a}, it was proposed to set $p_A = p_B = p$, thus making the probabilities identical and time-independent at the same time. While this assumption, along with the initial condition $I_0$, played a crucial role in estimating an upper bound of about $20\%$  for the number of agents with two names in their inventories~\cite{MARCHETTI2021} \footnote{The agents having two names in their inventories are often called bilinguals in the literature.}, it cannot be used to demonstrate the existence of fixed points A and B in full generality, as wrongly assumed in Ref.~ \cite{Marchetti-2020a}. In fact, consensus cannot be reached in such an initial setup due to the absence of spontaneous symmetry breaking~\cite{Goldstone1961, michel1980}~\footnote{Here the term  ``spontaneous'' indicates that the symmetry breaking occurs without the intervention of an external agent, similar to what happens with the Higgs mechanism in particle physics.}, similar to what happens  to the dynamical system associate with  MNG (see Section~\ref{introduction}).

Next, we shall demonstrate that the assumption $p_A = p_B = p$ cannot be valid throughout the entire time-evolution of the dynamical system associated with BNG. To illustrate this fact we computed 
cumulative distribution function (CDF) of the values taken by $p_A$ and $p_B$ during a single multi-agent simulation of a system with size $N=3,000$ (see Section~\ref{naming_games} for details).

\begin{figure}
\resizebox{0.50\textwidth}{!}{%
  \includegraphics{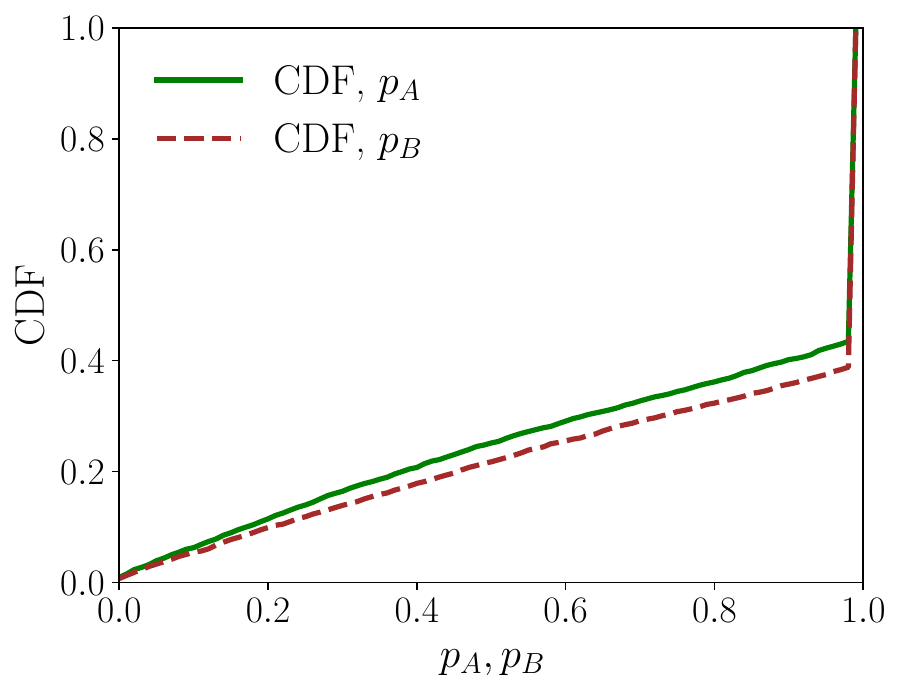}
}

\caption{The cumulative distribution function of $p_A$ (solid line) and $p_B$ (dashed line) as a function of the Bayesian probabilities $p_A$ and $p_B$, respectively, obtained from a simulation of a system with size $N=3,000$, starting from the initial condition $I_0$, and assuming the Erlang prior. The name learning thresholds for A and B,  were $n^{\ast}_{ex, A}=5$ and $n^{\ast}_{ex, B}=6$, respectively.  In such a case the consensus is reached at A.}
\label{fig:cdf}       
\end{figure}

In Fig.~\ref{fig:cdf}, the curves corresponding to CDF of $p_A$
(solid line) and $p_B$ (dashed line) are shown as functions of the Bayesian probabilities. From their different trends, it is clear that the assumption $p_A = p_B = p$ is untenable and may only be valid for a short time-interval. Furthermore, there is also a step-like progression toward unity in the two cumulative distribution functions as a direct consequence of the few-shot learning processes~\cite{Tenenbaum-1999}. The latter feature of this model demonstrates that it is implausible to consider such probabilities as smooth functions either of the number of positive examples or of time. Regarding their time-dependence, we cannot infer its exact functional form directly from the multi-agent simulations, due to their strong dependence on the system size $N$. However, Bayesian probabilities exhibit an evident monotonically increasing trend with time within a mean-field approximation~\cite{Marchetti-2020a}.

 With the above considerations in mind, we propose modeling both  $p_A\left(t\right)$ and $p_B\left(t\right)$  as a time-dependent logistic function (or sigmoid function)~\cite{Murphy-2012a}~\footnote{The logistic function is a common choice of activation function for the neural networks and deep learning.}. The logistic function has a  S-shaped  form that can be modified appropriately thanks to its two characteristic parameters: the logistic growth rate (or the steepness of the curve) $\eta$,  and the quantity $t_m$ that corresponds to the value of $t$  of function's midpoint. Denoting it by the symbol $g$, the logistic function reads:
\begin{equation}\label{eq:logistic}
g\left(t\right) = \frac{1}{1 + \exp\left(-\eta \left(t - t_m\right)\right)} \, .
\end{equation}

In the following, we shall give an example of how our probabilistic modeling yields reasonable results in accordance with the numerical simulations.
Without the loss of generality, we chose the following three values for  parameters $\eta$ and $t_m$: $\eta =2$, $t_{m, A}=10$ and $t_{m, B}=130$. Inserting $\eta =2, t_{m, A}=10$ and $\eta =2, t_{m, B}=130$ into Eq.~\ref{eq:logistic} we get two different logistic functions: $g_A$ and $g_B$, respectively. While such a value of $\eta$ guarantees that both $g_A$ and $g_B$ are step-like non-smooth functions of time $t$, the different  values of  $t_m$  i.e. $t_{m, A}$, $t_{m, B}$,  differentiate  $g_A$ from $g_B$. Furthermore, our  choice of parameters ensures that the learning of name A will be favored.

\begin{figure}

\resizebox{0.50\textwidth}{!}{%
  \includegraphics{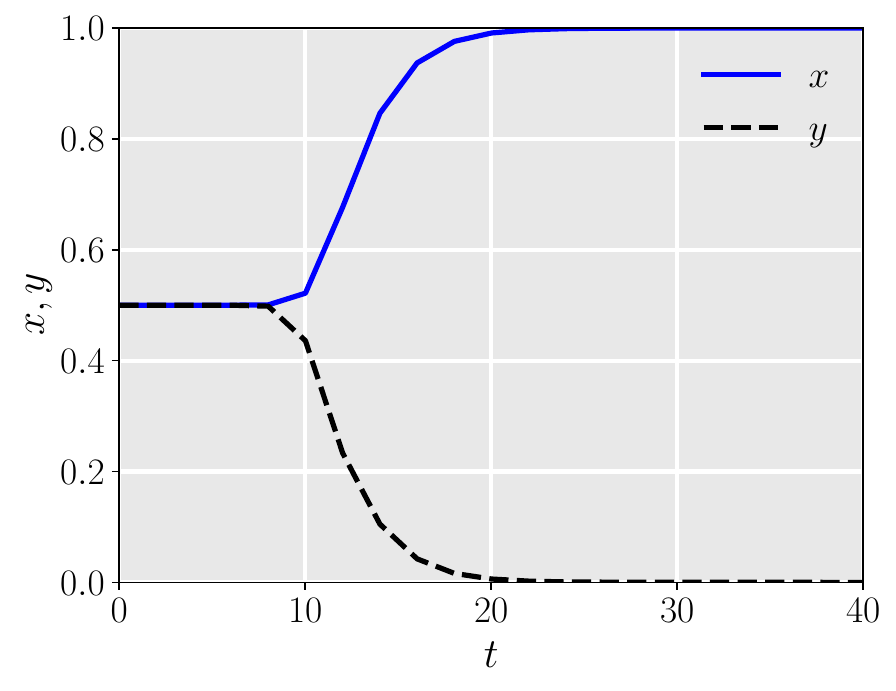}
}

\caption{Numerical solution  of Eqs.~\ref{eq:nonautonomous1}, \ref{eq:nonautonomous2} replacing the probabilities $p_A$ and $p_B$ with the logistic functions $g_A$ and $g_B$, respectively. The logistic function's parameters used are $\eta =2$ for both  $g_A$ and $g_B$, $t_{m, A}=10$ and $t_{m, B}=130$ for  $g_A$ and $g_B$, respectively.   The initial condition is  $I_0= \left(0.5, 0.5\right)$.  
 }
\label{fig:solutionBNG}       
\end{figure}

Next, we proceed to numerically integrate Eqs.~\ref{eq:nonautonomous1}, \ref{eq:nonautonomous2}, replacing $p_A$ with $g_A$ and $p_B$ with $g_B$, using the algorithm LSODA~\cite{petzold1983}, and assuming the initial condition $I_0= \left(0.5, 0.5\right)$. The numerical result shows that the solution $\left(x, y\right)$ asymptotically converges to $A = \left(1, 0\right)$ (see  Fig.~\ref{fig:solutionBNG}). Accordingly,  there exists now a spontaneous emergence of consensus  despite the dynamics following a mean-field deterministic prescription. Furthermore, by swapping the values of parameters $t_{m, A}$ and $t_{m, B}$, it is also straightforward to show that the solution converges to the other possible consensus state, that is,  $B= \left(0, 1\right)$.   These outcomes are attributed to the spontaneous symmetry breaking induced by the time-dependent probabilities, disrupting the invariance under the swap of variables $x$ and $y$ of Eqs.~\ref{eq:nonautonomous1} and ~\ref{eq:nonautonomous2}. Remarkably, the numerical integration  yields a solution $\left(x, y\right)$  that closely resembles that obtained from the numerical simulations with relatively large number of agents, i.e., $N  \gtrsim 200$ (see Ref.~\cite{Marchetti-2020a}).

We conclude by noting that the numerical solution $\left(x, y\right)$ splits into its two components $x$ and $y$ after a transient time during which $x=y$ (see Fig.~\ref{fig:solutionBNG}). Such an initial stage of dynamics,  can be thought as dictated by a dynamical system, whose differential equations  read: $\dot{x} = - p_B x y$, $\dot{y} = - p_A x y$. This observation clearly explains that in the initial stage of the dynamics, the time-evolution of the system is now primarily dictated by the Bayesian probabilities $p_A$ and $p_B$. Once again, it is their interplay and time-dependence that lead to the subsequent spontaneous symmetry-breaking process, essential for the emergence of consensus.

\section*{Acknowledgments}

The author would like to express gratitude to Steven Strogatz for his helpful insights on bifurcation, to Konstantinos "Kostas" Zygalakis for the important remarks about the effect of noise around critical points and bifurcations, and to Andrea Baronchelli for reading the manuscript.  The author extends thanks to Martti Raidal for the financial support via the European Regional Development Fund CoE program TK133 “The Dark Side of the Universe”, and to Els Heinsalu and Marco Patriarca for proposing the naming game as a research topic.

\appendix

\section{Sample Variance of the Random Variable $X_t$}\label{variance}

\begin{figure}

\resizebox{0.50\textwidth}{!}{%
  \includegraphics{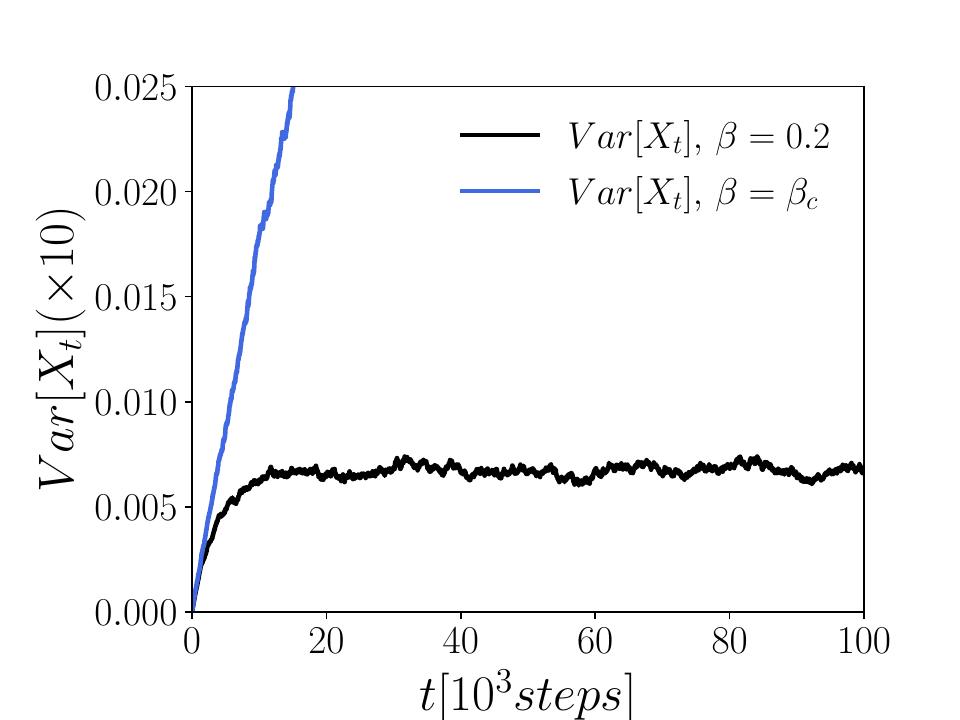}
}

\caption{Sample variance  $Var[X_t]$ of the random variable $X_t$  ($X_0 =0.5$) as function of time $t$. The variance is computed  from $1, 200$ trajectories of the $\beta$-NG model, assuming  $\beta=0.2$ (black line) and  $\beta=\beta_c$ (blue line), respectively. Note that for a better visualization we multiply  both sample variances by a factor $10$.}
\label{fig:sample_variance}      
\end{figure}

The theory of stochastic processes provide the following facts about the variance $Var[X_t]$ of Brownian motion and OU process. For the former it is expected that
the variance grows linearly with time, i.e.,  $Var[X_t] \propto  \sigma_1 t$ ($\sigma_1$ is a suitable positive parameter) 
while for the latter is given by the following equation~\cite{Oksendal2010}:
\begin{equation}\label{eq:variance_OU}
Var[X_t] = \frac{\sigma_{2}^{2}}{2\kappa} \left(1- e^{-2\kappa t}\right) 
 \, ,
\end{equation}
where $\sigma_2$ and $\kappa$ are two suitable parameters. In Fig.~\ref{fig:sample_variance} we plot the sample variances of the random variable $X_t$
as function of time $t$, computed  from $1, 200$ trajectories of the $\beta$-NG model, assuming  $\beta=0.2$ (black line) and  $\beta=\beta_c$ (blue line), respectively. It is evident that while the curve corresponding to $\beta=\beta_c$ exhibits the expected linear trend of a Brownian motion, the other one corresponding to $\beta=0.2$ correctly behaves according to Eq.~\ref{eq:variance_OU}. To quantitatively support our observations, we proceed to fit the data to the above formulas.
The curve fitting procedure by non-linear least squares yields the following optimal values for the parameters: $\sigma_1 = 0.00016 \pm 1.07 \times 10^{-8} $ with $R^{2} \approx 0.99$ and  $\sigma_2 = 0.013 \pm 7.7 \times 10^{-6} $,  $\kappa =  0.132 \pm 1.5 \times 10^{-4} $ with $R^{2} \approx 0.93$ where  $R^{2}$ denotes the coefficient of determination. These  values of $R^{2}$ indicate a good fit to the data, thereby further supporting our considerations regarding the type of stochastic processes for the problem at hand.

\bibliography{references}{}

\bibliographystyle{apsrev4-1}

\end{document}